\documentclass[prl,aps,twocolumn,superscriptaddress,showpacs]{revtex4}

\usepackage{pxfonts}
\usepackage{graphicx}
\begin{document}

\title{Kinetics of indirect excitons in the optically-induced trap}

\author{A.\,T. Hammack}
\author{L.\,V. Butov}
\affiliation{Department of Physics, University of California at San
Diego, La Jolla, CA 92093-0319}

\author{L. Mouchliadis}
\author{A.\,L. Ivanov}
\affiliation{Department of Physics and Astronomy, Cardiff
University, Cardiff CF24 3AA, United Kingdom}

\author{A.\,C. Gossard}
\affiliation{Materials Department, University of California at Santa
Barbara, Santa Barbara, California 93106-5050}

\begin{abstract}
We report on the kinetics of a low-temperature gas of indirect
excitons in the optically-induced exciton trap. The excitons in the
region of laser excitation are found to rapidly --- within
4\,ns --- cool to the lattice temperature $T=1.4$\,K, while the
excitons at the trap center are found to be cold --- essentially at the lattice
temperature --- even during the excitation pulse. The loading time of
excitons to the trap center is found to be about 40\,ns, longer than
the cooling time yet shorter than the lifetime of the indirect
excitons. The observed time hierarchy is favorable for creating a
dense and cold exciton gas in optically-induced traps and for
\emph{in situ} control of the gas by varying the excitation profile
in space and time before the excitons recombine.
\end{abstract}


\date{\today}

\maketitle

The modern advent of optical and magnetic trapping of atomic gases
\cite{Chu1998, Cohen-Tannoudji1998, Phillips1998, Ashkin2000} has
led to the observation of new states of matter such as Bose-Einstein
condensates (BEC) in atomic gases \cite{Cornell2002, Ketterle2002}
and atomic degenerate fermi-gases \cite{DeMarco1999}. Optical
trapping allowed the creation of a variety of potential reliefs for
atoms, including optical lattices \cite{Greiner2002,Chin2006}, and
control of these reliefs \emph{in situ}. One of the most important
characteristics, which is crucial for the observation, probe, and
control of these new states, is the trap loading time. Apparently,
studies of the degenerate gases in the traps require the fast
loading of the gases to the trap, preferably on a timescale less
than their lifetime in the trap. For the typical optical traps, the
loading times of the degenerate atomic gases are on the order of a
few tens of seconds while their lifetimes in the trap are on the
order of a few seconds \cite{Lewandowski2003,Streed2006}.

The system of excitons offers an opportunity to study degenerate
Bose-gases in trap potentials in semiconductor materials. Their low
mass results in a high temperature of quantum degeneracy for
excitons $T_{\rm dB}$, on the order of 1 K \cite{Keldysh1968}. For
instance, for excitons in a quantum well $T_{\rm dB} = 2\pi\hbar^2
n_{\rm 2D}/(mgk_{\rm B}) \simeq 3$\,K at the exciton density per
spin state $n_{\rm 2D}/g = 10^{10}$\,cm$^{-2}$. Exciton gases with
temperatures well below 1 K and densities higher than
$10^{10}$\,cm$^{-2}$ can be realized in coupled quantum well (CQW)
structures \cite{Butov2004}. The indirect excitons in CQWs can cool
down to these low temperatures due to their long lifetimes and high
cooling rates \cite{Butov2001}.

Indirect excitons in CQW can be confined by a variety of trapping
potentials including strain-induced traps
\cite{Wolfe1975,Trauernicht1983,Kash1988}, traps created by
laser-induced interdiffusion \cite{Brunner1992}, electrostatic traps
\cite{Zimmermann1997,Huber1998,Hammack2006a,Chen2006}, magnetic
traps \cite{Christianen1998}, and optically-induced traps
\cite{Hammack2006b}. Two of these trap types --- the electrostatic
and optically-induced traps --- show promise of rapid and effective
control of the excitons by varying in space and time the gate
voltage pattern and the laser intensity pattern, respectively.
Control of excitons by varying the electrostatic potential on a time
scale much shorter than the exciton lifetime was recently
demonstrated \cite{Winbow2007}. This indicates the feasibility of
studying excitons in controlled electrostatic traps. However, the
timescale of exciton control in an optically-induced trap remained
unknown before this work.

In this paper, we present the first studies of the spatial and
spectral kinetics of excitons in the optically-induced trap. The
results demonstrate a rapid loading of the trap by cold excitons on
a timescale less than the exciton lifetime and prove the feasibility
of accumulating a spatially confined dense and cold exciton gas as
well as the rapid \emph{in situ} control of excitons in the traps.

\begin{figure}
\begin{center}
\includegraphics[width=8.5cm]{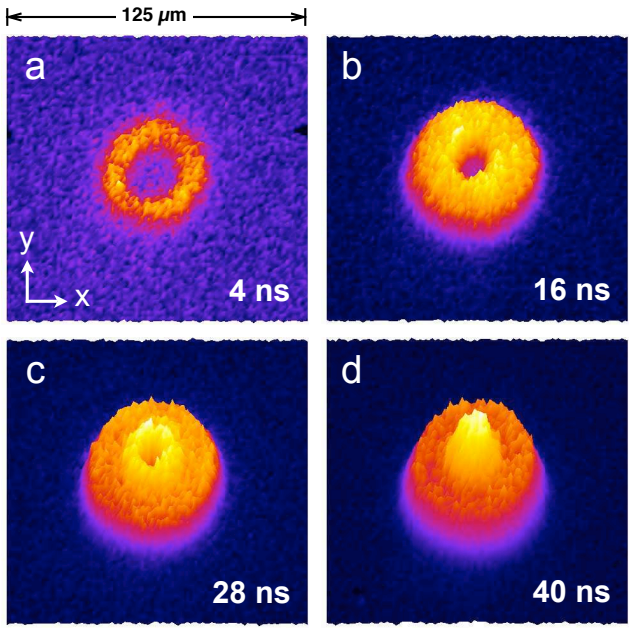}
\caption{Time resolved images of laser-induced trapping of excitons
collected by a photogated CCD. (a)-(d) $x$--$y$ plots of the PL
intensity from indirect excitons collected at delays of 4, 16, 28,
and 40\,ns relative to the start of 535\,nm pulsed laser diode
excitation in a 30\,$\mu$m diameter ring on the CQW sample. The
time-integration window for each image is 4\,ns. Average excitation
power $P_{\rm ex} = 75$\,$\mu$W. Bath temperature $T_{\rm b} =
1.4$\,K. $V_{\rm g} = 1.2$\,V.}
\end{center}
\end{figure}

The experimental measurements were performed using time-resolved
imaging with 4\,ns time-resolution and 2\,$\mu$m spatial resolution.
The excitons were photogenerated using rectangular laser excitation
pulses emitted by a pulsed semiconductor laser diode at 635\,nm. The
pulse duration was 500\,ns, the edge sharpness $< 1$\,ns, and the
repetition frequency 1\,MHz. The period and duty cycle were chosen
to provide ample time for the exciton gas to reach equilibrium
during the laser pulse and to allow complete decay of the indirect
exciton photoluminescence (PL) between the pulses. The pulses were
patterned into a laser excitation ring with a diameter of 30\,$\mu$m
and a ring thickness following a Gaussian profile of FWHM $= 2
\sigma \simeq 10$\,$\mu$m using a shadow mask. This ring-shaped laser
excitation created the in-plane spatial confinement of cold indirect
excitons by exploiting their mean-field repulsive interaction, which
forms the trapping profile \cite{Hammack2006b}. Time-dependent
spatial $x$--$y$ and spectral $Energy$--$y$ PL images were acquired
by a nitrogen-cooled CCD camera after passing through a time-gated
PicoStar HR TauTec intensifier with a time-integration window of
4\,ns. For $x$--$y$ imaging, an interference filter, chosen to match
the indirect exciton energy exclusively, is placed in the optical
path before the time-gated intensifier. In this manner, both the
high intensity low-energy bulk emission and direct exciton emission
are removed. The spectral filtering and time-gated imaging combine
to allow the direct visualization of the intensity profile of the
indirect exciton emission in spatial coordinates as a function of
delay time $\tau$ (see Fig.~1a-d). The time dependent measurements
of the indirect exciton spectra were acquired by placing the
time-gated intensifier and CCD after a single grating spectrometer.

The CQW structure used in these experiments was grown by molecular
beam epitaxy and contains two $8$\,nm GaAs QWs separated by a
$4$\,nm Al$_{0.33}$Ga$_{0.67}$As barrier (details on the CQW
structures can be found in \cite{Butov2004}). Indirect excitons in
the CQW structure are formed from electrons and holes confined to
different QWs. Separation between the electron and hole layers in
the CQW structure causes the optical lifetime $\tau_{\rm opt}$ of
the indirect excitons to exceed that of regular direct excitons by
orders of magnitude and is $\tau_{\rm opt} \simeq 50$\,ns at $V_{\rm
g} = 1.2$\,V for the studied CQW sample. The exciton density profile
was estimated from the measured exciton energy profiles in $E$--$y$
coordinates using the relation between the energy and density of the
indirect excitons $\delta E = u_0 n_{\rm 2D}$, where $u_0 = 4\pi e^2
d / \varepsilon_{\rm b}$
\cite{Yoshioka1990,Zhu1995,Ben-Tabou2001,Ivanov2002}. For the
studied CQW sample $\varepsilon_{\rm b} = 12.9$ and $d = 11.5$\,nm
\cite{Butov2004}.

\begin{figure}
\begin{center}
\includegraphics[width=8.5cm]{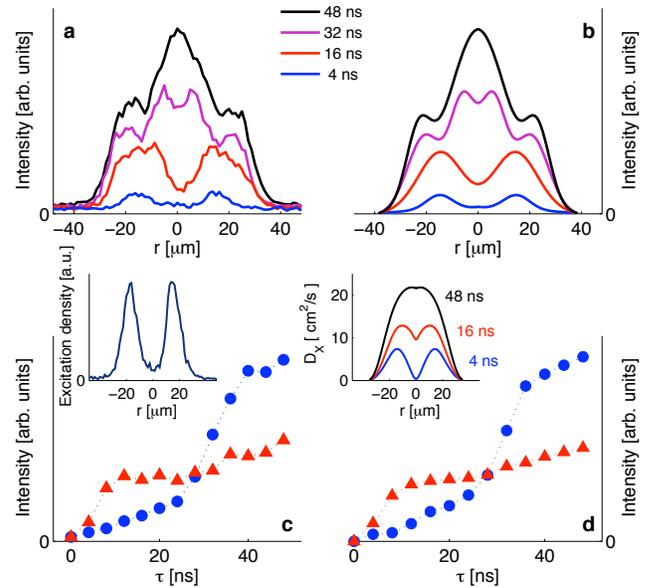}
\caption{Kinetics of the indirect exciton PL profile following the
onset of the ring-shaped laser excitation pulse. The measured (a)
and calculated (b) cross-sections of the indirect exciton PL across
the diameter of the laser excitation ring as a function of time. The
measured (c) and calculated (d) indirect exciton PL intensity at the
ring center (blue circles) and in the area of the laser excitation
ring (red triangles) as a function of time. The time-integration
window for each profile (a,b) and point (c,d) is 4\,ns. $\tau=0$ and
$\tau=500$\,ns correspond to the onset and termination of the
rectangular laser excitation pulse, respectively. Left inset: The
ring-shaped laser excitation profile. Right inset: The calculated
radial dependence of the exciton diffusion coefficient for different
time delays. Average excitation power $P_{\rm ex} = 75\, \mu$W. Bath
temperature $T_{\rm b} = 1.4$\,K. $V_{\rm g} = 1.2$\,V. }
\end{center}
\end{figure}

The kinetics presented in Fig.\,2c-d demonstrate that the exciton
pattern reaches a stationary state and, therefore, the trap loading
is completed within about 40\,ns. Thus the trap loading time is
about the lifetime of indirect excitons in this CQW sample (see
Fig.\,3c). Note that in another CQW sample with a larger separation
between the electron and hole layers the lifetime of indirect
excitons reaches several microseconds \cite{Winbow2007}, which is
much larger than the trap loading time reported here. 
However, trap loading on a time scale comparable to the lifetime in the trap is already
favorable for studies of confined degenerate gases
\cite{Lewandowski2003,Streed2006}.

Furthermore, for the excitation power used in the experiments, the
density of excitons at the trap center reaches $n_{\rm 2D} \simeq
1.4\times10^{10}$\,cm$^{-2}$ within 40\,ns. As shown below, the
exciton gas is cold, essentially at the lattice temperature, within
the area $\sim 100 \mu$m$^2$ around the trap center. Therefore,
$\sim 10^4$ cold excitons are loaded to the trap over the course of
40\,ns. The corresponding collection rate of cold excitons to the
optically-induced exciton trap exceeds $10^{11}$ excitons/second.

In order to model the experimental data we use the coupled
drift-diffusion and thermalization equations for the exciton
effective temperature $T$ and density $n_{\rm 2D}$:
\begin{eqnarray}
\! \! \! \! \! \! \! \! \! \! \! \! \frac{\partial n_{\rm
2D}}{\partial t} &=& \nabla \big[ D_{\rm x} \nabla n_{\rm 2D} +
\mu_{\rm x} n_{\rm 2D} \nabla ( u_0 n_{\rm 2D}) \big] - {n_{\rm
2D} \over \tau_{\rm opt}} + \Lambda ,
\label{diff} \\
\frac{\partial T}{\partial t} &=& \left( \frac{\partial T}{\partial
t} \right)_{n_{\rm 2D}} + \ S_{\rm pump} + S_{\rm opt} \,,
\label{therm}
\end{eqnarray}
where $D_{\rm x}$, $\mu_{\rm x}$, $\tau_{\rm opt} = \tau_{\rm
opt}(n_{\rm 2D},T)$ and $\Lambda = \Lambda(r_{\|},t)$ are the
diffusion coefficient, mobility, optical lifetime and generation
rate of indirect excitons, respectively
\cite{Ivanov2002,Ivanov2006}. The term $(\partial T/\partial
t)_{n_{\rm 2D}}$ describes cooling of indirect excitons due to
their coupling with bulk LA-phonons \cite{Ivanov1999}, and the
terms $S_{\rm pump}$ and $S_{\rm opt}$ refer to heating of
indirect excitons by the laser field and recombination heating or
cooling of the particles, as detailed in Ref.\,\cite{Ivanov2006}.
The stationary solution of these equations for laser induced traps
is discussed in Ref. \cite{Ivanov2006}. In the present work,
Eqs.\,(\ref{diff})-(\ref{therm}) are analyzed to model the
trapping kinetics, i.e., in the space and time domain.

The effective $n_{\rm 2D}$-dependent screening of the disorder
potential $U_{\rm rand}({\bf r}_{\|})$ by dipole-dipole interacting
indirect excitons is crucial for the drastic decrease of the time
needed to fill up the optically-induced trap with indirect excitons.
The thermionic model, derived from Eq.\,(\ref{diff}) for a
long-range correlated disorder potential, yields the effective
diffusion coefficient:
\begin{equation}
D_{\rm x} = D_{\rm x}^{(0)} \exp \bigg[ - \frac{U^{(0)}}{k_{\rm B}T
+ u_0 n_{\rm 2D}} \bigg] \,, \label{coeff}
\end{equation}
where $U^{(0)} = 2 \langle |U_{\rm rand}({\bf r}_{\|}) - \langle
U_{\rm rand}({\bf r}_{\|}) \rangle|\rangle$ and $D_{\rm x}^{(0)}$ is
the in-plane diffusion coefficient in the absence of CQW disorder
\cite{Ivanov2002}.

\begin{figure}
\begin{center}
\includegraphics[width=8.5cm]{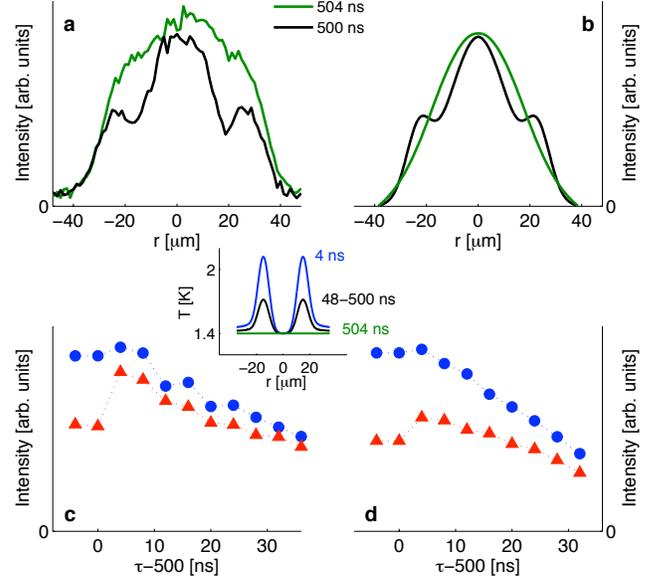}
\caption{Indirect exciton PL following the termination of the
ring-shaped laser excitation pulse. The measured (a) and calculated
(b) cross-sections of the indirect exciton PL across the diameter of
the laser excitation ring as a function of time. The measured (c)
and calculated (d) indirect exciton PL intensity at the ring center
(blue circles) and in the area of the laser excitation ring (red
triangles) as a function of time. The time-integration window for
each profile (a,b) and point (c,d) is 4\,ns. $\tau=0$ and
$\tau=500$\,ns correspond to the onset and termination of the
rectangular laser excitation pulse, respectively. Inset: The
calculated temperature across the ring diameter in the beginning of
the laser excitation pulse at $\tau=4$\,ns (blue), in the stationary
regime achieved during the 500~ns-long excitation pulse about 40 ns
after its start $\tau=48$ to $500$\,ns (black), and 4 ns after the
termination of the laser excitation pulse at $\tau=504$\,ns (green).
Average excitation power $P_{\rm ex} = 75\, \mu$W. Bath temperature
$T_{\rm b} = 1.4$\,K. $V_{\rm g} = 1.2$\,V. }
\end{center}
\end{figure}

By using Eq.\,(\ref{coeff}), the time $\tau_{\rm trav}$ required for
an indirect exciton to travel from the boundary of an annular trap
of radius $R$ to its center is estimated as
\begin{eqnarray}
\tau_{\rm trav}^{(1)} &=& \frac{R^2}{D_{\rm x}^{(0)}} \,
e^{U^{(0)}/k_{\rm B}T} \ \ \ \ \mbox{for} \ \ k_{\rm
B} T \gg u_0n_{\rm 2D}\, , \label{t1} \\
\tau_{\rm trav}^{(2)} &=& \frac{R^2}{D_{\rm x}^{(0)}} \,
\frac{k_{\rm B}T}{u_0 n_{\rm 2D}^{(0)}} \, e^{U^{(0)}/(u_0 n_{\rm
2D}^{(0)})} \ \ \ \ \mbox{for} \ \ k_{\rm B} T \ll u_0n_{\rm 2D}\,
, \label{t2}
\end{eqnarray}
where $n_{\rm 2D}^{(0)}$ is the density of photoexcited indirect
excitons at the boundary of the trap, and the first, low-density
limit ($n_{\rm 2D} < 10^9$\,cm$^{-2}$) refers to the unscreened
disorder potential, while the second, high-density limit ($n_{\rm
2D} \geq 10^{10}$\,cm$^{-2}$) deals with effective mean-field
screening of $U_{\rm rand}({\bf r}_{\|})$. In the low-density
limit, the excitons are essentially localized by disorder, the
diffusion coefficient is small, $D_{\rm x} = D_{\rm x}^{(0)}
\exp[-U^{(0)}/(k_{\rm B} T)] \sim 0.1$\,cm$^2$/s, and the in-plane
transport of excitons out of the excitation spot cannot be seen
because in this case $\tau_{\rm trav}^{(1)} \gg \tau_{\rm opt}$.
In contrast, for $k_{\rm B} T \ll u_0n_{\rm 2D}$ the diffusion
coefficient is large, $D_{\rm x} = D_{\rm x}^{(0)}
\exp[-U^{(0)}/(u_0n_{\rm 2D})] \sim 10$\,cm$^2$/s, giving rise to
the drastic decrease of the characteristic travel time $\tau_{\rm
trav}$: According to Eqs.\,(\ref{t1})-(\ref{t2}), $\tau_{\rm
trav}^{(2)} = \beta \tau_{\rm trav}^{(1)} \ll \tau_{\rm
trav}^{(1)}$ with the dimensionless smallness parameter $\beta
\sim 10^{-3}-10^{-5}$ (for $T \sim 1$\,K, $U_0 \sim 1$\,meV, and
$n_{\rm 2D} \sim 10^{10}$\,cm$^{-2}$). The transition from
localized to delocalized indirect excitons is indeed observed with
increasing density \cite{Butov2002,Ivanov2006,Hammack2006b}.

The calculated change of the exciton diffusion coefficient, due to
screening of CQW disorder, is shown in the right inset of Fig.\,2. A
drastic increase of $D_{\rm x}$ with increasing $n_{\rm 2D}$ is
consistent with the above estimates. For our experiments,
evaluations with Eq.\,(\ref{t2}) yield $\tau_{\rm trav}^{(2)} \simeq
4.6$\,ns against $\tau_{\rm opt} \simeq 50$\,ns, i.e., the condition
$\tau_{\rm trav}^{(2)} \ll \tau_{\rm opt}$ is clearly met.

The numerical simulations were done by using control parameters
consistent with those found in our previous studies
\cite{Butov2001,Hammack2006b} of similar structures: $u_0 = 1.6
\times 10^{-10}$\,meVcm$^2$, $U^{(0)} = 1.2$\,meV, $D_{\rm
x}^{(0)} = 35$\,cm$^2$/s, $m = 0.215\,m_0$ ($m_0$ is the free
electron mass), $E_{\rm x} = 1.55$\,eV, and the deformation
potential of exciton -- LA-phonon interaction $D_{\rm DP} =
6.5$\,eV. As can be seen in Figs.~2 and 3, the measured and
calculated kinetics of the exciton spatial patterns are in
quantitative agreement.

When the laser pulse is switched off, a jump in the exciton PL is
observed in the region of laser excitation (Fig.\,3). As shown in
the previous studies \cite{Feldmann1987}, only low-energy excitons
with the kinetic energy from the radiative zone, $E < E_0 \simeq
E_{\rm x}^2\varepsilon_{\rm b}/(2mc^2)$, are optically active. The
PL-jump is caused by the increase of the radiative zone
population, due to the cooling of high-energy optically-dark
excitons after abruptly switching off the laser excitation
\cite{Butov2001}. The time-resolved imaging experiments show that
the PL-jump is observed only in the region of laser excitation,
where indirect excitons are effectively heated by the laser field.
In this region, after switching off the laser excitation
(Fig.\,3), the excitons rapidly cool down to the lattice
temperature $T_{\rm b}=1.4$\,K within\,4 ns, the time resolution
of the current experiments. The calculated cooling (i.e.
thermalization) time is $\tau_{\rm th} \simeq 0.2\,\mbox{ns}$.
Both the experiment and calculations show that the exciton cooling
time to the lattice temperature is much shorter than the exciton
lifetime $\tau_{\rm opt} \simeq 50$ ns.

More importantly, the data show no PL-jump at the trap center
(Fig.\,3). This proves that the excitons at the trap center are
cold, essentially at the lattice temperature $T_{\rm b}$, even in
the presence of the excitation pulse. The numerical simulations of
the exciton temperature profile, $T = T(r_{\|})$, plotted in the
inset of Fig.\,3 for various time delays, are consistent with this
observation. The hierarchy of times, $\tau_{\rm th} \ll \tau_{\rm
trav} \ll \tau_{\rm opt}$, results in complete thermalization of
indirect excitons during their travel from the boundary of the
optically-induced trap to its center, where heating from the laser
excitation is negligible.

In conclusion, we have found that the characteristic loading time
of an optically-induced trap with cold excitons is on the scale of
tens nanoseconds. The observed hierarchy of times (exciton cooling
time$) < ($trap loading time$) < ($exciton lifetime in the trap)
is favorable for creating a dense and cold exciton gas in the
optically-induced traps and its \emph{in situ} control by varying
the excitation profile in space and time before the exciton
recombination.

This work is supported by ARO grant W911NF-05-1-0527 and NSF grant
DMR-0606543.


\begin{thebibliography}{26}

\bibitem{Chu1998}
S.~Chu, Rev. Mod. Phys. {\bf 70}, 685 (1998).

\bibitem{Cohen-Tannoudji1998}
C.~N. Cohen-Tannoudji, Rev. Mod. Phys. {\bf 70}, 707 (1998).

\bibitem{Phillips1998}
W.~D. Phillips, Rev. Mod. Phys. {\bf 70}, 721 (1998).

\bibitem{Ashkin2000}
A.~Ashkin, IEEE Journal on Selected Items in Quantum Electronics
{\bf  6}, 841 (2000).

\bibitem{Cornell2002}
E.~A. Cornell, C.~E. Wieman, Rev. Mod. Phys. {\bf 74}, 875 (2002).

\bibitem{Ketterle2002}
W.~Ketterle, Rev. Mod. Phys. {\bf 74}, 1131 (2002).

\bibitem{DeMarco1999}
B.~DeMarco, D.~S. Jin, Science {\bf 285}, 1703 (1999).

\bibitem{Greiner2002}
M.~Greiner, O.~Mandel, T.~Esslinger, T.~W. H\"ansch, and I.~Bloch,
Nature {\bf 415}, 39 (2002).

\bibitem{Chin2006}
J.~K. Chin, D.~E. Miller, Y.~Liu, C.~Stan, W.~Setiawan, C.~Sanner,
K.~Xu, and W.~Ketterle, Nature {\bf 443}, 961 (2006).

\bibitem{Lewandowski2003}
H.~J. Lewandowski, D.~M. Harber, D.~L. Whitaker, and E.~A. Cornell,
J. Low Temp. Phys. {\bf 132}, 309 (2003).

\bibitem{Streed2006}
E.~W. Streed, A.~P. Chikkatur, T.~L. Gustavson, M.~Boyd, T. Torii,
D.~Schneble, G.~K. Campbell, D.~E. Pritchard, and W.~Ketterle, Rev.
of Sci. Inst. {\bf 77}, 023106 (2006).

\bibitem{Keldysh1968}
L.~V. Keldysh and A.~N. Kozlov, Zh. Eksp. Teor. Fiz. {\bf 54}, 978
(1968) [Sov. Phys. JETP {\bf 27}, 521 (1968)].

\bibitem{Butov2004}
L.~V. Butov, J. Phys.: Condens. Matter {\bf 16}, R1577 (2004).

\bibitem{Butov2001}
L.~V. Butov, A.~L. Ivanov, A.~Imamoglu, P.~B. Littlewood, A.~A.
Shashkin, V.~T. Dolgopolov, K.~L. Campman, and A.~C. Gossard, Phys.
Rev. Lett. {\bf 86}, 5608 (2001).

\bibitem{Wolfe1975}
J.~P. Wolfe, W.~L. Hansen, E.~E. Haller, R.~S. Markiewicz,
C.~Kittel, and C.~D. Jeffries, Phys. Rev. Lett. {\bf 34}, 1292
(1975).

\bibitem{Trauernicht1983}
D.~P. Trauernicht,  A.~Mysyrowicz,  and J.~P. Wolfe, Phys. Rev. B
{\bf 28}, 3590 (1983).

\bibitem{Kash1988}
K.~Kash, J.~M. Worlock, M.~D. Sturge, P.~Grabbe, J.~P. Harbison,
A.~Scherer, and P.~S.~D. Lin, Appl. Phys. Lett. {\bf 53}, 782
(1988).

\bibitem{Brunner1992}
K.~Brunner, U.~Bockelmann, G.~Abstreiter, M.~Walther, G.~B{\"{o}}hm,
G.~Tr{\"{a}}nkle, and G.~Weimann, Phys. Rev. Lett. {\bf 69}, 3216
(1992).

\bibitem{Zimmermann1997}
S.~Zimmermann, A.~O. Govorov, W.~Hansen, J.~P. Kotthaus, M.~Bichler,
and W.~Wegscheider, Phys. Rev. B {\bf 56}, 13414 (1997).

\bibitem{Huber1998}
T.~Huber, A.~Zrenner, W.~Wegscheider, and M.~Bichler, Phys. Stat.
Sol. (a) {\bf 166}, R5 (1998).

\bibitem{Hammack2006a}
A.~T. Hammack, N.~A. Gippius, Sen Yang, G.~O. Andreev, L.~V. Butov,
M.~Hanson, A.~C. Gossard, J. Appl. Phys. {\bf 99}, 066104 (2006).

\bibitem{Chen2006}
G.~Chen, R.~Rapaport, L.~N. Pffeifer, K.~West, P.~M. Platzman,
S.~Simon, Z.~V{\" o}r{\" o}s, D.~Snoke, Phys. Rev. B {\bf 74},
045309 (2006).

\bibitem{Christianen1998}
P.~C.~M. Christianen, F.~Piazza, J.~G.~S. Lok, J.~C. Maan, and
W.~van~der Vleuten, Physica B {\bf 249}, 624 (1998).

\bibitem{Hammack2006b}
A.~T. Hammack, M.~Griswold, L.~V. Butov, L.~E. Smallwood, A.~L.
Ivanov, A.~C. Gossard, Phys. Rev. Lett. {\bf  96}, 227402 (2006).

\bibitem{Winbow2007}
A.~G. Winbow, A.~T. Hammack, L.~V. Butov, A.~C. Gossard, Nano
Letters {\bf 7}, 1349 (2007).

\bibitem{Yoshioka1990}
D.~Yoshioka, A.~H. MacDonald, J. Phys. Soc. Jpn. {\bf 59}, 4211
(1990).

\bibitem{Zhu1995}
X.~Zhu, P.~B. Littlewood, M.~Hybertsen, T.~Rice, Phys. Rev. Lett.
{\bf 74}, 1633 (1995).

\bibitem{Ben-Tabou2001}
S.~Ben-Tabou de-Leon, B.~Laikhtman, Phys. Rev. B {\bf 63}, 125306
(2001).

\bibitem{Ivanov2002}
A.~L. Ivanov, Europhys. Lett. {\bf 59}, 586 (2002).

\bibitem{Ivanov2006}
A.~L. Ivanov, L.~E. Smallwood, A.~T. Hammack, Sen Yang, L.~V. Butov,
and A.~C. Gossard, Europhys. Lett. {\bf 73}, 920 (2006).

\bibitem{Ivanov1999}
A.~L. Ivanov, P.~B. Littlewood, and H.~Haug, Phys. Rev. B {\bf 59},
5032 (1999).

\bibitem{Butov2002}
L.~V. Butov, A.~C. Gossard, and D.~S. Chemla, Nature {\bf 418}, 751
(2002).

\bibitem{Feldmann1987}
J.~Feldmann, G.~Peter, E.~O. Gobel, P. Dawson, K. Moore, C. Foxon,
R.~J. Elliott, Phys. Rev. Lett. {\bf 59}, 2337 (1987).

\end{thebibliography}
\end{document}